\documentclass[a4paperl,preprint]{revtex4}
\usepackage{ae}
\usepackage{physics}
\usepackage[T1]{fontenc}
\usepackage[ansinew]{inputenc}
\usepackage{amsmath}
\usepackage{amssymb}
\usepackage{amsthm}
\usepackage[caption=false]{subfig}
\usepackage{multirow}
\usepackage{array}
\usepackage{wrapfig}
\usepackage{epstopdf}
\usepackage{xparse}
\usepackage{dcolumn}
\usepackage[mathscr]{euscript}

\usepackage[]{graphicx}

\usepackage{color}
\usepackage[colorlinks]{hyperref}
\usepackage{lscape}
\theoremstyle{remark}

\graphicspath{ {./images1/} }
\usepackage{tikz}
\usepackage{textcomp}
\begin{document}
\title{Measurement dependence can enhance security in a quantum network}
\author{Amit Kundu}
\email{amit8967@gmail.com}
\affiliation{Department of Applied Mathematics, University of Calcutta, Kolkata- 700009, India}
\author{Debasis Sarkar}
\email{dsarkar1x@gmail.com}
\affiliation{Department of Applied Mathematics, University of Calcutta, Kolkata- 700009, India}

\begin{abstract}
 Network Nonlocality is an advanced study of quantum nonlocality that comprises network structure beyond Bell's theorem. The development of quantum networks has the potential to bring a lot of technological applications in sevaral quantum information processing tasks.  Here, we are focusing on how the role of the independence of the measurement choices of the end parties in a network works and  can be used to enhance the security in a quantum network. In both three-parties two-sources bilocal network and four-parties three-sources star network scenarios, we are able to show, a practical way to understand the relaxation of the assumptions to enhance a real security protocol if someone wants to breach in a network communications. Theoratically, we have proved that by relaxing the independence of the measurement choices of only one end party we can create a Standard Network Nonlocality(SNN) and more stronger Full Network Nonlocality(FNN) and we can get maximum quantum violation by the classical no-signalling local model. We are able to distinguish between two types of network nonlocality in the sense that the FNN is stronger than SNN, i.e., FNN states all the sources in a network need to distribute nonlocal resources. 
\end{abstract}

\date{\today}
\maketitle
\section{Introduction}
A network consists of several number of parties connected in many possible ways carrying a correlation between them. This network structure is very important in quantum information processing. Nowadays, the urge to understand quantum correlations in network scenarios has reached much attention and development\cite{network1, network2, network3}. There is a lot of interest in the case where several quantum sources share quantum correlations between many connected parties in various ways and reveal quantum nonlocality\cite{swap, review,triA,noinbi,silva23,neural}. 
The crucial assumption for a quantum network is the independence of the several quantum sources, along with the assumptions of realism, locality and measurement independence, used to create Bell-type inequalities\cite{bellpaper}. In a typical Bell scenario, if a correlation from a quantum source between two parties, say, Alice and Bob, violates the celebrated Bell-CHSH inequality, then the correlation is non-local in nature. By this non-local correlation, we mean that the correlation from the quantum source cannot be simulated by a local deterministic model if those three assumptions are intact. There are ways\cite{banik3, banik4, banik5} to classically reproduce nonlocality by relaxing some of the assumptions like measurement independence. In\cite{banik12, banik4, Hall1, Hall2}, they simulated a classical model that violates Bell-CHSH inequality by relaxing measurement independence. They made a scenario where the choices of the input of the parties, Alice and Bob, are not fully random. The hidden variables control the input choices and outcomes. They also showed that 30\% and 59\% experimental measurement dependence is needed to simulate quantum correlation for both-sided and one-sided dependency, respectively. 

Now, what about the measurement independence assumption in a network structure. Till date we have no significant results. Here, in this letter, we are focusing on how this assumption uncovers novel forms of quantum nonlocal correlations inherent to the network structure. How one can characterizes the forms of non-locality and explore their potential for applications in quantum information processing by applying measurement dependence. Further, what if we loose some of the assumptions and introduce measurement dependence for the end parties in a network structure and what will be the status of the correlation remain in the given network structure? The results that we are going to explore are very important in the sense that they reveal new physics in a network which will help us to understand the structure of the correlation better and to apply it in practical implications, we name it Boss Breaches, where the middle party has a lot of power to control the whole network correlation just by influencing the one end party, asking to relax the randomness of the choice of measurement. By controlling the correlation we mean that Bob can simulate classically the whole quantum maximal correlation distributed in the networks with his classical variable selector and no one can know. This is true for both Bilocal network and star network.
\\ In short, we will first describe the Network Nonlocality, specially Standard Network Nonlocality and Full network Nonlocality in section II. In section III, we will show how we are imposing mathematically measurement dependence to the end party in a network. In the next two section we will show that we can reach the maximum quantum violation of in Bilocal and Star network by classical model and its practical aspects respectively. We will end this literature with the conclusion.
\section{Network Nonlocality}
Consider a network composed of $m$ number of sources with $k$ number of parties having input choices $x_k$ and output $a_k$. The correlations maintained in this network will satisfy a network local model if it obeys;
\begin{equation}
p(a_1...a_k|x_1...x_k)=\idotsint d\lambda_{1}\rho_{1}(\lambda_{1})... d\lambda_{m}\rho_{m}(\lambda_{m})\prod_i p(a_{i}|x_{i},{\bar{\lambda}}_{i})
\end{equation}
where $\lambda_{i}$'s are the local variables associated with each source, and $\rho_i$'s are the probability densities, and $\bar{\lambda_i}$'s are the set of local hidden variables of the sources connected with the $i'{th}$ party, otherwise the correlation will be called non n-local. This scenario will reduce to the typical Bell scenario if  $m=1$ and $k=2$.\\
In this letter, we will consider first the trivial network scenario called the bilocal scenario when $m=2$, $k=3$ and then a star network scenario with $m = 3$, $k = 4$\cite{nlocal}. For the bilocal scenario, probability distribution decomposed as,
$$p(a_1,a_2,a_3|x_1,x_2,x_3)=\int\int d\lambda_{1}\rho_{1}(\lambda_{1})d\lambda_{2}\rho_{2}(\lambda_{2})p(a_{1}|x_{1},\lambda_1)p(a_{2}|x_{2},\lambda_1,\lambda_2)p(a_{3}|x_{3},\lambda_3)$$ 
can be called bilocal in nature and by violating
\begin{equation}\label{BIin}
\sqrt{I} + \sqrt{J} \leq 1,
\end{equation}
 we define a correlation to be non-bilocal(equ:\ref{I1}, \ref{J1} for $I$ and $J$). In\cite{gisin17}, they showed if only one source is Bell nonlocal, then also the whole correlation over the bilocal network will violate bilocal inequality. Later in\cite{Kundu20, andreoli}, they generalised it in the n-local scenario. In recent times, A. Pozas- Krestjens et. al. introduced a new idea called Full Network Nonlocality(FNN) where by creating the maximum quantum violation of the bilocal inequality with a local and PR box correlation shared between three parties in a bilocal scenario, they concluded that the inequality(equ:\ref{BIin}) only confirms Standard Network Nonlocality(SNN). But in the trivial star network scenario with three sources and four parties they showed that the correlation will provide $S = \sqrt{2}$, where, generally for n-local scenario,
\begin{equation}\label{3local}
S_n = \sum_{j=1}^{2^{n-1}} |I_j|^{\frac{1}{n}} \leq 2^{n-2}
\end{equation}  
with $I_j = \frac{1}{2^n}\sum_{x_1...x_n}(-1)^{g_j(x_1...x_n)}\langle A_{x_1}^1 ... A_{x_n}^n B^j\rangle$\cite{nlocal,Kundu20}, coming from three maximally entangled states shared from three different sources(fig:\ref{nlocal}) is Full Network Nonlocal correlation, i.e., it is not reproducible by making any of the three sources local. \\
\section{Imposing Measurement Dependence}
Now, relaxing the assumption of measurement independence, we will show that under certain amount of measurement dependence, only one party measurement dependence can classically simulate not only maximal quantum correlation in bilocal scenario(SNN) but also maximal quantum correlation in star  shaped 3-local scenario(FNN). Both are defined by the maximal violation of the equ:\ref{BIin} and equ:\ref{3local}. This also allows us to use this maximal correlation in various information protocol tasks by controlling the randomness of the measurement choices. For the more complex quantum network structure the dependence of measurement choices is increasing in very interesting way with the complexity of the structure which we will discuss later.

Consider the scenario with three parties sharing two sources of independent hidden states; this is the simplest case where our assumption of independent 
sources with hidden states make sense. In such tripartite scenario, Bell's locality assumption reads $$P(a,b,c|x,y,z)= \int d\lambda \rho(\lambda)P(a|x, \lambda)P(b|y, \lambda)P(c|z, \lambda).$$
Here given the hidden state $\lambda$, the outputs $a$, $b$ and $c$ of the three parties, for inputs $x$, $y$ and $z$ are determined, respectively, by the local distributions 
$P(a|x, \lambda)$, $P(b|y, \lambda)$ and $P(c|z, \lambda)$. The hidden states $\lambda$ follow the distribution $\rho(\lambda)$, normalized such that $\int \rho(\lambda)d\lambda$=1.
If we assume that the response of the three parties depends only on the states $\lambda_{1}$ or $\lambda_{2}$ characterizing the systems that they receive from the sources $S_{1}$ or $S_{2}$, respectively, we write
$$P(a,b,c|x,y,z)= \int \int d\lambda_{1} d\lambda_{2} \rho(\lambda_{1}, \lambda_{2})P(a|x, \lambda_{1})P(b|y, \lambda_{1}, \lambda_{2})P(c|z, \lambda_{2}).$$
This is a Bilocal correlation and it will jump to the tripartite Bell correlation if $\lambda_{1}$ = $\lambda_{2}$ = $\lambda$.
The crucial assumption for this correlation is that the two sources are independent to each other which is covered by the condition $$\rho(\lambda_{1}, \lambda_{2})=\rho(\lambda_{1})\rho(\lambda_{2})$$
with following distributions $\int \rho(\lambda_{1})d\lambda_{1}$ = $\int \rho(\lambda_{2})d\lambda_{2}$ = 1.
\begin{figure}[h]
\centering
\includegraphics[width=12cm, height=9cm]{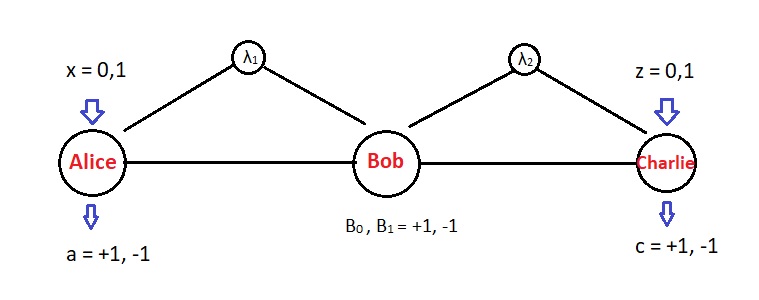}
\caption{Bilocal Network scenario}
\label{biloc}
\end{figure}

In \cite{cava11,cava12}, the first non-linear inequalities that allow one to efficiently capture 2-local correlations were derived. Here we focus on an inequality 
presented in \cite{cava12}, which we will refer to as the Bilocal inequality(equ:\ref{BIin}). Consider that Alice and Charlie receive binary inputs, $x$, $x'$ and $z$, $z'$ and must provide binary outputs, denoted by
$a$ = $\pm$1 and $c$ = $\pm$1, respectively. The middle party Bob always performs the same measurements with four possible outcomes, the usual BSM(Bell State Measurements). Denote Bob's outcome by two bits $B_{0}$ = $\pm$1 and $B_{1}$ = $\pm$1. The Bilocal Inequality reads 
$S_{biloc} \equiv \sqrt{I} + \sqrt{J} \le 1$ where
\begin{equation}\label{I1}
I \equiv \frac{1}{4}\langle(A_{x}+A_{x'})B_{0}(C_{z}+C_{z'})\rangle
\end{equation}

\begin{equation}\label{J1}
J \equiv \frac{1}{4}\langle(A_{x}-A_{x'})B_{1}(C_{z}-C_{z'})\rangle
\end{equation}

A non-bilocal correlation will violate this inequality. But after that, no signalling correlations also form a polytope, bounded by $\max(|I|,|J|)\le1$; 
In particular, a no-signalling correlation such that $a+b+c=xy+yz$(mod 2), can easily be realised with two PR boxes, reaches the value $I$ = $J$ =1.
We now introduced here the degree of measurement dependence for this scenario. The dependence means that the hidden variable distribution controls the measurement choices and it is quantified by the variational
distance between the distribution of the shared parameter for any pair of measurement settings. Here we impose one-sided measurement dependence, i.e., on Alice's side or Charlie's side, as the bilocal scenario carries a structural symmetry. 
We define the degree of measurement dependence, $M_1 = \int |\rho(\lambda_1|x) - \rho(\lambda_1|x')$ and $M_2 = \int |\rho(\lambda_2|z) - \rho(\lambda_2|z')$ for Alice's side and Charlie's side respectively. 
Clearly the values of $M_1$ and $M_2$ can vary from 0 to 2. The experimental measurement independence, i.e., the fraction of measurement independence corresponding
to a given model is defined as; $F_1/(F_2) =( 1 - \frac{M_1}{2})/(1 - \frac{M_2}{2})$. Applying those conditions to the non linear Bilocal inequality, we have the modified inequality, 
$S_{biloc} \equiv \sqrt{I} + \sqrt{J} \le 1 + \sqrt{\frac{M_1}{2}}$ or $S_{biloc} \equiv \sqrt{I} + \sqrt{J} \le 1 + \sqrt{\frac{M_2}{2}}$ for Alice's part or Charlie's part respectively. See apendix A for the proof.
\section{Classical model for SNN and FNN}
The largest known quantum violation by the bilocal inequality is $S_{biloc} = \sqrt{2}$ and it is obtained by each source emitting singlet states $|\psi^-\rangle$ = $(|01\rangle - |10\rangle)/\sqrt{2}$. In a no-signaling deterministic model, the one-sided degree of measurement dependence that is required to reproduce the correlation corresponding to the value of the maximum quantum violation is $M_1 = 2(\sqrt{2}-1)^2$, and the experimental measurement independence $F_1 = 2\sqrt{2} - 2 = 82\%$. Here measurement dependence $M_1 = 2$ and $F_1 = 0$ provides the value of the inequality from the P-R correlation where no experimental independence is there. Here we provide a no signalling deterministic model that reproduces the above correlation and also compatible with the results.
\begin{center}
\begin{tabular}{|c|c|c|c  c  c  c|c c|c c|}
\hline
B & $|B|$ & $\lambda_1$ & X  &  X' &  Z &  Z' & $\rho(\lambda_1|\lambda_2,X,Z)$ & $\rho(\lambda_1|\lambda_2,X,Z')$ & $\rho(\lambda_1|\lambda_2,X',Z)$ & $\rho(\lambda_1|\lambda_2,X',Z')$ \\
\hline
\multirow{3}{4em}{$B_0$} & 00 & $\lambda_{1,1}$ & -a & -a & -a & -a & 0 & 0 & p & p\\
& 01 & $\lambda_{1,2}$ & b & b & b & b & 1 & 1 & 1-p & 1-p\\
\hline
\multirow{3}{4em}{$B_1$} & 10 & $\lambda_{1,1}$ & a & a & -a & a & 0 & 0 & p & p\\
& 11 & $\lambda_{1,2}$ & b & b & b & -b & 1 & 1 & 1-p & 1-p\\
\hline
\end{tabular}
\end{center}
In the table we describe the singlet correlation in a bilocal network with measurement dependency only on the Alice's side with the outcome containing the variables $\lambda_{1,1}$ and $\lambda_{1,2}$ for the measurement setting $X$ and $X'$. We define $B_0$, $B_1$ as a selector of hidden variable($\lambda's$) distribution. $a$ and $b$ can pick any value $\pm1$. The probability distribution is defined by a single parameter $0\leq p\leq 1$ here. With some simple calculation, we can get the value of $I = 4$ and $J = 4p$; this violates the Bilocal inequality $S_{biloc}$ by $\sqrt{p}$. So the degree of measurement dependence is $M_1 = 2p$ which clearly shows that $0\leq M_1\leq 2$ and $0\leq F_1 \leq 1$. Therefore, for $S_{biloc} = 1 - \sqrt{\frac{M_1}{2}}$, the amount of one-sided measurement independence is $F = 1 + \sqrt{p}$ in agreement with the analytical measurement independence of $82\%$. Here one can also see the maximum Bilocal violation(PR correlation) can be simulated by only sacrificing the full measurement independence of one party($p = 1$). 
\begin{figure}[h]
\centering
\includegraphics[width=12cm, height=7cm]{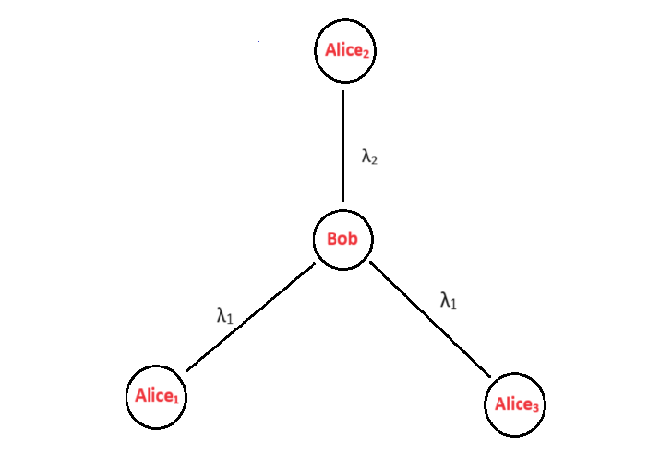}
\caption{3-local scenario}
\label{nlocal}
\end{figure}
\\We are focusing now on the trivial star network scenario with three sources and four parties connected illustrated in the fig:\ref{nlocal}. Here the end parties, $Alices$, are linked to a single part $Bob$ who performs a fixed measurement with an outcome represented by a string $b = b_1b_2b_3 \in (0,1)^3$. For this scenario we can get a inequality from equ:\ref{3local}, which is $$\sqrt[3]{|I_1|} + \sqrt[3]{|I_2|} +\sqrt[3]{|I_3|} + \sqrt[3]{|I_4|} \leq 2$$. Now maximally entangled quantum states from three sources violates this inequality with the value of $2\sqrt{2}$. From \cite{fullNN}, we can say that quantum correlation reaching $2\sqrt{2}$ are Full Network Nonlocal. We are going to show that this Full Network Nonlocal correlation can be simulated by no-signalling classical local model just by imposing $92\%$ measurement dependence on only one party. With similar approach of bilocal inequality, we have the modified inequality for four parties star network by imposing any of that $Alices$ and it would provide us(Appendix B),

\begin{equation}
\sqrt[3]{|I_1|} + \sqrt[3]{|I_2|} +\sqrt[3]{|I_3|} + \sqrt[3]{|I_4|} \leq 2 + \sqrt[3]{4M_1}
\end{equation}
In a no-signaling deterministic model, the degree of measurement dependence for one party that is required to reproduce the correlation corresponding to the value of the maximum quantum violation is $M_1 = 2(\sqrt{2}-1)^3$, and the experimental measurement independence $F_1 = 1 - \frac{M_1}{2} = 92\%$. Here we provide a no signalling deterministic model that reproduces the above correlation and also compatible with the results.
\begin{center}
\begin{tabular}{|c|c|c|c c c c c c|c c c c|c c c c|}
\hline
B & $|B|$ & $\lambda_1$ & X  &  X' & Y & Y' & Z &  Z' & $\rho_{\lambda1|\lambda2,1}$ & $\rho_{\lambda1|\lambda2,2}$ & $\rho_{\lambda1|\lambda2,3}$ & $\rho_{\lambda1|\lambda2,4}$ & $\rho_{\lambda1|\lambda2,1'}$ & $\rho_{\lambda1|\lambda2,2'}$ & $\rho_{\lambda1|\lambda2,3'}$ & $\rho_{\lambda1|\lambda2,4'}$ \\
\hline
\multirow{3}{4em}{$B^1$} & 000 & $\lambda_{1,1}$ & a & a & a & a & a & a & 0 & 0 & 0 & 0 & p & p & p & p\\
& 001 & $\lambda_{1,2}$ & b & b & b & b & b & b & 1 & 1 & 1 & 1 & 1-p & 1-p & 1-p & 1-p\\
\hline
\multirow{3}{4em}{$B^2$} & 010 & $\lambda_{1,1}$ & a & a & -a & a & a & a & 0 & 0 & 0 & 0 & p & p & p & p\\
& 011 & $\lambda_{1,2}$ & b & -b & b & -b & b & b & 1 & 1 & 1 & 1 & 1-p & 1-p & 1-p & 1-p\\
\hline
\multirow{3}{4em}{$B^3$} & 100 & $\lambda_{1,1}$ & a & -a & -a & -a & -a & a & 0 & 0 & 0 & 0 & p & p & p & p\\
& 101 & $\lambda_{1,2}$ & b & b & b & b & b & -b & 1 & 1 & 1 & 1 & 1-p & 1-p & 1-p & 1-p\\
\hline
\multirow{3}{4em}{$B^4$} & 110 & $\lambda_{1,1}$ & a & a & a & -a & a & -a & 0 & 0 & 0 & 0 & p & p & p & p\\
& 111 & $\lambda_{1,2}$ & -b & b & b & -b & b & -b & 1 & 1 & 1 & 1 & 1-p & 1-p & 1-p & 1-p\\
\hline
\end{tabular}
\end{center}
Here, $\rho_{\lambda1|\lambda2,1}$, $\rho_{\lambda1|\lambda2,2}$, $\rho_{\lambda1|\lambda2,3}$ and $\rho_{\lambda1|\lambda2,4}$ are written in short form of $\rho_{\lambda_1|\lambda_2,\lambda_3,X,Y,Z}$, $\rho_{\lambda_1|\lambda_2,\lambda_3,X,Y,Z'}$, $\rho_{\lambda_1|\lambda_2,\lambda_3,X,Y',Z}$ and $\rho_{\lambda_1|\lambda_2,\lambda_3,X,Y',Z'}$ respectively and similarly for $\rho_{\lambda1|\lambda2,1'}$, $\rho_{\lambda1|\lambda2,2'}$, $\rho_{\lambda1|\lambda2,3'}$ and $\rho_{\lambda1|\lambda2,4'}$, they are $\rho_{\lambda_1|\lambda_2,\lambda_3,X',Y,Z}$, $\rho_{\lambda_1|\lambda_2,\lambda_3,X',Y,Z'}$, $\rho_{\lambda_1|\lambda_2,\lambda_3,X',Y',Z}$ and $\rho_{\lambda_1|\lambda_2,\lambda_3,X',Y',Z'}$ respectively.

\begin{figure}
\centering
\includegraphics[width=12cm, height=7cm]{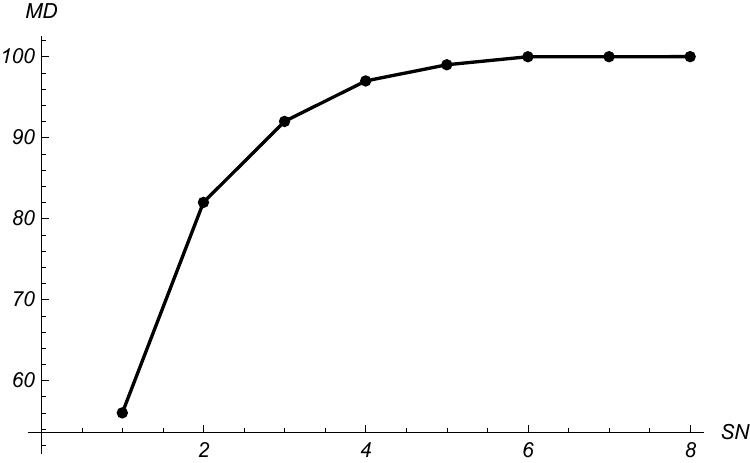}
\caption{Percentage of Measurement Dependence vs Sources Numbers.}
\label{SNMD}
\end{figure}

In this table we describe the maximal quantum correlation in a four parties star network with measurement dependency only on one of the Alices side with the outcome containing the variables $\lambda_{1,1}$ and $\lambda_{1,2}$ for the measurement setting $X$ and $X'$. Here similarly we define $B^1$, $B^2$, $B^3$ and $B^4$ as a selector of hidden variable($\lambda's$) distribution and $a$, $b$ pick the value $1$. The same simple calculation provides the value of $\sqrt[3]{|I_1|} + \sqrt[3]{|I_2|} +\sqrt[3]{|I_3|} + \sqrt[3]{|I_4|} \leq 2 + 2\sqrt[3]{p}$; this violates the quantity $S_3$(equ:\ref{3local}) by $2\sqrt[3]{p}$. So the degree of measurement dependence is $M_1 = 2p$ which clearly shows that $0\leq M_1\leq 2$ and $0\leq F_1 \leq 1$. Therefore, for $S_3 = 2 + \sqrt[3]{\frac{M_1}{2}}$, the amount of one-sided measurement independence is $F = 1 - p = 1 - (\sqrt{2} - 1)^3$ in agreement with the analytical measurement independence of $92\%$. Similarly with bilocal scenario, one can also see the maximum 3-local violation(PR correlation) can be simulated by only sacrificing the full measurement independence of one party($p = 1$). 
Now we generalize the idea of imposing measurement dependence for further complex network structure with increasing source number in a star network and fig:\ref{SNMD} shows the need of measurement dependence for only one party to simulate the maximal quantum violation for each structure with increasing source number. When the source number is greater than five, nearly full dependence is needed because of the structural complexity.

\begin{figure}
\centering
\includegraphics[width=14cm, height=6cm]{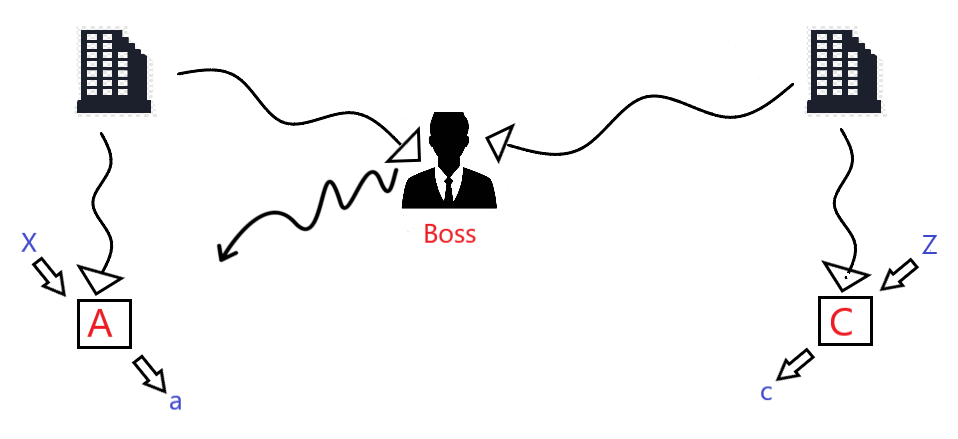}

\caption{Schematic diagram of a bilocal network in a practical way. Here middle party is a $Boss$, who is powerful here and $A$ and $C$ are the two parties. The two buildings are the sources of quantum states. The wavy line from $Boss$ to $A$ represents the influence that $A$ got from $Boss$ to relax the measurement independence.}
\label{real}
\end{figure}

\begin{figure}
\centering
\includegraphics[width=14cm, height=6cm]{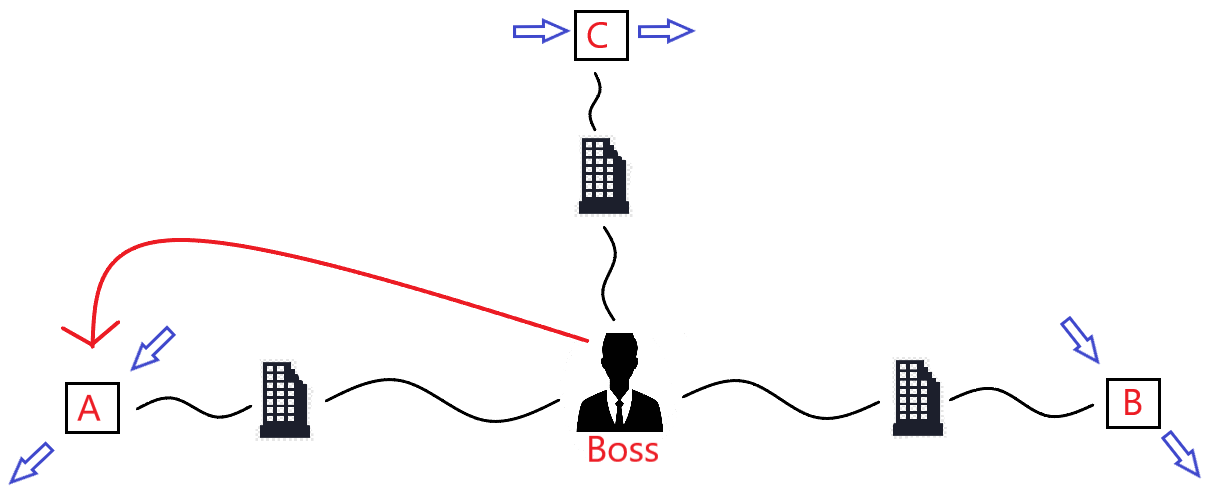}

\caption{This is the  3-local network in a practical way. Here also middle party is a $Boss$, who is powerful here and $A$, $B$ and $C$ are the three parties. The sources are three. Here the red line from $Boss$ to $A$ represents the influence.}
\label{realnet}
\end{figure}
\section{Practical Aspects}
Thus in the network nonlocality scenario, the assumption of measurement independence is very strong. This work can be implemeted in a practical way of security breaches. We name it $Boss$ $Breaches$. Fig:(\ref{real}) and fig:(\ref{realnet}) represent a schematic diagram of a practical way of describing a network with multiple parties and multiple sources. Here the buildings represent companies which are the two quantum sources, distributing quantum states between A, Boss and C in the preferred way. Now we are thinking of a situation where the Boss is a corrupt person and he is the only one who connects A and C in a security protocol through quantum sources and they are running any specific security protocol based on the maximum quantum violation of Bilocal Inequality(\ref{BIin}). Now the perfect picture is where A and C make measurements on their particle received from two different companies randomly and the Boss is neutral and makes a single input with four output measurements and with both purely quantum sources the correlation reaches the maximum value of the inequality. Now think of a corrupt situation where the corrupt Boss wants to run the same security protocol with the same maximum violation of the inequality but this time only A is with the Boss and A has a special power to relax the measurement randomness and will do the same at the Boss's order. Now our results corresponding to the Bilocal scenario confirm that whenever Boss wants to twist the security he asks A to relax the measurement randomness and asks him to relax it to $82\%$ so that the Boss can simulate classically, with his own local variable, the whole correlation(SNN) which violates the Bilocal Inequality maximally and C, the Companies, even A do not know that the violation is coming from the quantum states or the classical simulation. This is what we can say about a security fallacy where only one party just listens to the Boss and the Boss can control the whole correlation classically without any other person's knowledge.

Now this situation can also happen in a stronger correlation structure. If we consider the scenario with three parties connected again with the corrupt Boss and the companies are distributing quantum states to the parties and Boss (fig:\ref{realnet}), the relaxation of $92\%$ on the measurement randomness on only one party can give the Boss power to simulate classically the whole maximum quantum correlation(FNN) in a 3-local scenario again without the knowledge of the other. The classical model scripted in the tables can certify this statement. This is a practically rich way to understand the network theory and the relaxation of the measurement independence of any one party. 
We can think of another way to use this theory and it is randomness certification. Using a typical Bell scenario randomness certification has been done many times but in a multi-party multi-sources quantum network scenario, the assumption of measurement dependence can be used for randomness certification in a more complicated manner. 

\section{Conclusion}

This work captures that if any security protocol is running based on the maximal quantum violation of network inequality for the Bilocal scenario and star network scenario, then the middle party has the strength to activate his local classical model without informing anyone and still can get the maximal violation just by asking one end party not to choose the measurement inputs randomly. This strong statement we can make here just because we showed here how effective it is if we can lose some amount of measurement independence in the quantum network scenario that we can simulate, classically the network nonlocal correlation. In the trivial quantum network scenario, Bilocal Network, Network Nonlocality comes by violating the Bilocal non-linear inequality when the two sources are non-local in nature\cite{cava12}. According to the\cite{gisin17, andreoli, Kundu20}, the nature of the two sources can vary,  which is true for n-locality also. In\cite{fullNN}, by producing maximum quantum violation for a slightly complex network scenario, a three-source star network, by one local and one N-S correlation and they framed the network nonlocality in two classes, Standard Network Nonlocality(SNN) and Full Network Nonlocality(FNN). The former correlation is the correlation which can come from a situation where any one source can be the local source, but the latter is the correlation which confirms that all sources are nonlocal in nature. This work shows that in a bilocal scenario, the SNN can be revealed by a no-signalling classical model with 82$\%$ measurement dependence on only one party. Whereas the FNN, the correlation in the three-source four-party scenario when all the sources produce nonlocal states, can also be simulated by no-signalling classical model with 92$\%$ measurement dependence on only one party. This is a very strong result and to the best of our knowledge, this is the first work of finding optimal experimental dependence in network structure for both SNN and FNN. We here also incorporate the result of how the measurement dependence on only one party depends on the number of sources in a star network scenario. Simulating n-locality with $n$ sources network will be great work to do in future. This work is solely a step forward to better understand the correlation of SNN and FNN and is also very important to proceed with the idea in the more complex and non-trivial network structure.

  Recently in\cite{I} the effect of relaxing the assumption of source independence in network scenario are discussed. An interesting extension of this work can be the study of the joint effect of source dependence and measurement dependence in a quantum network scenario.
\section{Acknowledgements}
 The authors AK and DS acknowledge the work as part of QUest initiatives by DST India.

\onecolumngrid
\pagebreak
\appendix
\section{modified inequality with measurement dependence assumption for Bilocal scenario}\label{AppBilocal}
 The factorizable form of the $I$ and $J$ from the Bilocal inequality;
\begin{eqnarray}
I =\frac{1}{4} | \int \int d\lambda_{1}d\lambda_{2} \rho(\lambda_{1}, \lambda_{2})V(A_{x}|\lambda_{1})V(B_{0})V(C_{z}|\lambda_{2})+V(A_{x}|\lambda_{1})V(B_{0})V(C_{z'}|\lambda_{2})\nonumber \\
+V(A_{x'}|\lambda_{1})V(B_{0})V(C_{z}|\lambda_{2})+V(A_{x'}|\lambda_{1})V(B_{0})V(C_{z'}|\lambda_{2})|\nonumber\\ \label{eq:one}
\end{eqnarray}

\begin{eqnarray}
J = \frac{1}{4}| \int \int d\lambda_{1}d\lambda_{2} \rho(\lambda_{1}, \lambda_{2})V(A_{x}|\lambda_{1})V(B_{1})V(C_{z}|\lambda_{2})-V(A_{x}|\lambda_{1})V(B_{1})V(C_{z'}|\lambda_{2}) \nonumber\\
-V(A_{x'}|\lambda_{1})V(B_{1})V(C_{z}|\lambda_{2})+V(A_{x'}|\lambda_{1})V(B_{1})V(C_{z'}|\lambda_{2})| \nonumber\\ \label{eq:two}
\end{eqnarray}
Here for given value of $\lambda(\lambda_{1}, \lambda_{2})$ each observable takes a definite value, i.e.,
$$V(A_{x}|\lambda_{1}), V(B_{0}|\lambda_{1}\lambda_{2}), V(C_{z}|\lambda_{2})$$
and 
$$V(A_{x'}|\lambda_{1}), V(B_{1}|\lambda_{1}\lambda_{2}), V(C_{z'}|\lambda_{2}).$$
With fixed value ${+1, -1}$. Now from equ:[\ref{eq:one} and \ref{eq:two}] we get;
\begin{eqnarray}
	I\leq\frac{1}{4}\int|[\rho(\lambda_1|x)V(A_x)V(B_0)V(C_z)\rho(\lambda_2|z)+\rho(\lambda_1|x)V(A_x)V(B_0)V(C_{z'})\rho(\lambda_2|z')  \nonumber\\
	+\rho(\lambda_1|x')V(A_{x'})V(B_0)V(C_z)\rho(\lambda_2|z)+\rho(\lambda_1|x')V(A_{x'})V(B_0)V(C_{z'})\rho(\lambda_2|z')]| \nonumber\\ \label{eq:three}
\end{eqnarray}
 and
 \begin{eqnarray}
J\leq\frac{1}{4}\int|[\rho(\lambda_1|x)V(A_x)V(B_1)V(C_z)\rho(\lambda_2|z)-\rho(\lambda_1|x)V(A_x)V(B_1)V(C_{z'})\rho(\lambda_2|z') \nonumber\\
-\rho(\lambda_1|x')V(A_{x'})V(B_1)V(C_z)\rho(\lambda_2|z)+\rho(\lambda_1|x')V(A_{x'})V(B_1)V(C_{z'})\rho(\lambda_2|z')]|\nonumber\\ \label{eq:four}
\end{eqnarray}
as the absolute value of an integral is less than or equal to an integral of the absolute value of the integrand($|\int f(x)dx|\leq\int|f(x)|dx$). From equ:[\ref{eq:three}] and equ:[\ref{eq:four}], taking out the density function, we get
$$I\leq\frac{1}{4}\int|\rho(\lambda_1|x)[V(A_x)V(B_0)V(C_z)\rho(\lambda_2|z)+V(A_x)V(B_0)V(C_{z'})\rho(\lambda_2|z')]$$ 
$$+\rho(\lambda_1|x')[V(A_{x'})V(B_0)V(C_z)\rho(\lambda_2|z)+V(A_{x'})V(B_0)V(C_{z'})\rho(\lambda_2|z')]|$$ and 
$$J\leq\frac{1}{4}\int|\rho(\lambda_1|x)[V(A_X)V(B_1)V(C_z)\rho(\lambda_2|z)-V(A_x)V(B_1)V(C_{z'})\rho(\lambda_2|z')]$$ 
$$+\rho(\lambda_1|x')[V(A_{x'})V(B_1)V(C_z')\rho(\lambda_2|z')-V(A_{x'})V(B_1)V(C_{z})\rho(\lambda_2|z)]|.$$  It goes as,
$$I\leq\frac{1}{4}\int|\rho(\lambda_1|x)V(A_x)V(B_0)V(C_z)[\rho(\lambda_2|z)+\frac{V(C_{z'})}{V(C_z)}\rho(\lambda_2|z')]$$ 
$$+\rho(\lambda_1|x')V(A_{x'})V(B_0)V(C_z)[\rho(\lambda_2|z)+\frac{V(C_{z'})}{V(C_z)}\rho(\lambda_2|z')]|$$ and 
$$J\leq\frac{1}{4}\int|\rho(\lambda_1|x)V(A_x)V(B_1)V(C_z)[\rho(\lambda_2|z)-\frac{V(C_{z'})}{V(C_z)}\rho(\lambda_2|z')]$$ 
$$+\rho(\lambda_1|x')[V(A_{x'})V(B_1)V(C_{z'})[\frac{V(C_{z'})}{V(C_z)}\rho(\lambda_2|z')-\rho(\lambda_2|z)]|.$$
Since Alice's and Bob's observables attains a value $\pm1$, we can write,
$$I\leq\frac{1}{4}\int|[\rho(\lambda_1|X) + k'\rho(\lambda_1|X')][\rho(\lambda_2|Z) + k\rho(\lambda_2|Z')]|$$ and
$$J\leq\frac{1}{4}\int|[\rho(\lambda_1|X) - k'\rho(\lambda_1|X')][\rho(\lambda_2|Z) - k\rho(\lambda_2|Z')]|,$$ where $k=\frac{V(C_{z'})}{V(C_z)}$ and $k'=\frac{V(A_{x'}}{V(A_x)}$ takes a value $+1$ or $-1$ for a given hidden variable. We can see the symmetry of the integrand of $I$ and $J$ won't effect the changing of the value of $k$ and $k'$. So by changing the value of each $k$ and $k'$ we get;
$$4 + M_1M_2$$ and 
$$M_1M_2 + 4.$$
We here only consider one-sided measurement dependence so we can use $M_1 = 2$ or $M_2 = 2$.
So the $S_{biloc} = \sqrt{I} + \sqrt{J} \leq 1 + \sqrt{\frac{M_1}{2}}$ for Alice part of $1 + \sqrt{\frac{M_2}{2}}$ for Bob part.

\section{modified inequality for star network scenario}

For trivial star network scenario: Consider trivial star network for three sources and four party. The 3-local inequality is, $$S_3 = \sum_{j=1}^{4} |I_j|^{\frac{1}{3}} \leq 2$$
where 
$$I_1 = \frac{1}{8}(\langle A^1_XA^2_XA^3_XB^1\rangle + \langle A^1_XA^2_XA^3_{X'}B^1\rangle + \langle A^1_XA^2_{X'}A^3_XB^1\rangle + \langle A^1_XA^2_{X'}A^3_{X'}B^1\rangle +$$
 $$\langle A^1_{X'}A^2_XA^3_XB^1\rangle + \langle A^1_{X'}A^2_XA^3_{X'}B^1\rangle + \langle A^1_{X'}A^2_{X'}A^3_XB^1\rangle + \langle A^1_{X'}A^2_{X'}A^3_{X'}B^1\rangle)$$

$$I_2 = \frac{1}{8}(\langle A^1_XA^2_XA^3_XB^2\rangle + \langle A^1_XA^2_XA^3_{X'}B^2\rangle - \langle A^1_XA^2_{X'}A^3_XB^2\rangle - \langle A^1_XA^2_{X'}A^3_{X'}B^2\rangle $$ 
$$-\langle A^1_{X'}A^2_XA^3_XB^2\rangle - \langle A^1_{X'}A^2_XA^3_{X'}B^2\rangle + \langle A^1_{X'}A^2_{X'}A^3_XB^2\rangle + \langle A^1_{X'}A^2_{X'}A^3_{X'}B^2\rangle)$$

$$I_3 = \frac{1}{8}(\langle A^1_XA^2_XA^3_XB^3\rangle - \langle A^1_XA^2_XA^3_{X'}B^3\rangle + \langle A^1_XA^2_{X'}A^3_XB^3\rangle - \langle A^1_XA^2_{X'}A^3_{X'}B^3\rangle $$ 
$$- \langle A^1_{X'}A^2_XA^3_XB^3\rangle + \langle A^1_{X'}A^2_XA^3_{X'}B^3\rangle - \langle A^1_{X'}A^2_{X'}A^3_XB^3\rangle + \langle A^1_{X'}A^2_{X'}A^3_{X'}B^3\rangle)$$

$$I_4 = \frac{1}{8}(\langle A^1_XA^2_XA^3_XB^4\rangle - \langle A^1_XA^2_XA^3_{X'}B^4\rangle - \langle A^1_XA^2_{X'}A^3_XB^4\rangle + \langle A^1_XA^2_{X'}A^3_{X'}B^4\rangle $$ 
$$ +\langle A^1_{X'}A^2_XA^3_XB^4\rangle - \langle A^1_{X'}A^2_XA^3_{X'}B^4\rangle - \langle A^1_{X'}A^2_{X'}A^3_XB^4\rangle + \langle A^1_{X'}A^2_{X'}A^3_{X'}B^4\rangle)$$

Using similar approach we get,
$$I_1 \leq \frac{1}{8}\int[\rho(\lambda_1|A_X^1) + k\rho(\lambda_1|A_{X'}^1)][\rho(\lambda_2|A_X^2) + k\rho(\lambda_2|A_{X'}^2)][\rho(\lambda_3|A_X^3) + k\rho(\lambda_3|A_{X'}^3)]$$

$$I_2 \leq \frac{1}{8}\int[\rho(\lambda_1|A_X^1) + k\rho(\lambda_1|A_{X'}^1)][\rho(\lambda_2|A_X^2) - k\rho(\lambda_2|A_{X'}^2)][\rho(\lambda_3|A_X^3) - k\rho(\lambda_3|A_{X'}^3)]$$

$$I_3 \leq \frac{1}{8}\int[\rho(\lambda_1|A_X^1) - k\rho(\lambda_1|A_{X'}^1)][\rho(\lambda_2|A_X^2) + k\rho(\lambda_2|A_{X'}^2)][\rho(\lambda_3|A_X^3) - k\rho(\lambda_3|A_{X'}^3)]$$

$$I_3 \leq \frac{1}{8}\int[\rho(\lambda_1|A_X^1) - k\rho(\lambda_1|A_{X'}^1)][\rho(\lambda_2|A_X^2) - k\rho(\lambda_2|A_{X'}^2)][\rho(\lambda_3|A_X^3) + k\rho(\lambda_3|A_{X'}^3)]$$

Where $k$ signifies the same quantity which for a given local variable either takes +1 or -1.
For every possible value of $k$ we can have,
$$\sqrt[3]{|I_1|} + \sqrt[3]{|I_2|} +\sqrt[3]{|I_3|} + \sqrt[3]{|I_4|} \leq 1 + \sqrt[3]{\frac{2M_2M_3}{8}} + \sqrt[3]{\frac{M_12M_3}{8}} + \sqrt[3]{\frac{M_1M_22}{8}}
$$
For any one sided measurement dependence, we can get,
$$\sqrt[3]{|I_1|} + \sqrt[3]{|I_2|} +\sqrt[3]{|I_3|} + \sqrt[3]{|I_4|} \leq 2 + \sqrt[3]{4M_1},$$ by considering only one party dependence.
\end{document}